\documentclass{WileyMSP-template}
\usepackage{graphicx}
\usepackage{dcolumn}
\usepackage{bm}

\usepackage[utf8]{inputenc}
\usepackage[T1]{fontenc}
\usepackage{mathptmx}
\usepackage{etoolbox}
\usepackage{xcolor}
\usepackage{physics}
\usepackage{amsmath}
\usepackage{amssymb}
\usepackage{multirow}
\usepackage{graphicx}
\usepackage{float}
\usepackage{hyperref}
\renewcommand{\hbar}{\mathchar'26\mkern-9mu h}

\begin{document}

\pagestyle{fancy}

\title{Giant anomalous Hall and Nernst conductivities in magnetic all-$d$ metal Heusler alloys}

\maketitle


\author{Md Farhan Tanzim$\dag$}
\author{Nuno Fortunato$\dag$}
\author{Ilias Samathrakis}
\author{Ruiwen Xie*}
\author{Ingo Opahle}
\author{Oliver Gutfleisch}
\author{Hongbin Zhang}

$\dag${These authors contributed equally to this work}

\dedication{}
\begin{affiliations}
	
Md Farhan Tanzim, Nuno Fortunato, Ilias Samathrakis, Ingo Opahle, Hongbin Zhang\\
Theory of Magnetic Materials, Institute of Materials Science, Technical University of Darmstadt, 64287 Darmstadt, Germany.

Oliver Gutfleisch \\
Functional Materials, Institute of Materials Science, Technical University of Darmstadt, 64287 Darmstadt, Germany

Ruiwen Xie\\
Theory of Magnetic Materials, Institute of Materials Science, Technical University of Darmstadt, 64287 Darmstadt, Germany. \\
Email Address: ruiwen.xie@tmm.tu-darmstadt.de
\end{affiliations}


\keywords{All-\textit{d} Heusler, Transport properties, High-throughput}

\begin{abstract}

All-\textit{d} Heuslers are a category of novel compounds combining versatile functionalities such as  caloric responses and spintronics with enhanced mechanical properties.
Despite the promising transport properties (anomalous Hall (AHC) and anomalous Nernst (ANC) conductivities) shown in the conventional Co$_2$XY Heuslers with \textit{p}-\textit{d} hybridization, the all-\textit{d} Heuslers with only \textit{d}-\textit{d} hybridization open a new horizon to search for new candidates with outstanding transport properties. In this work, we evaluate the AHC and ANC for thermodynamically stable ferro/ferri-magnetic all-$d$-metal regular Heusler compounds based on high-throughput first-principles calculations.
It is observed that quite a few materials exhibit giant AHCs and ANCs, such as cubic Re$_2$TaMn with an AHC of 2011 S/cm, and tetragonal Pt$_2$CrRh with an AHC of 1966 S/cm and an ANC of 7.50 A/mK. Comprehensive analysis on the electronic structure reveals that the high AHC can be attributed to the occurrence of the Weyl nodes or gapped nodal lines in the neighbourhood of the Fermi level. The correlations between such transport properties and the number of valence electrons are also thoroughly investigated, which provides a practical guidance to tailor AHC and ANC via chemical doping for transverse thermoelectric applications.

\end{abstract}


\section{\label{sec:intro}Introduction}
In the last two decades, topological phenomena driven by the nontrivial geometric phase~\cite{xiao2010berry}  and topological materials have attracted significant attention; they are especially promising for sophisticated electronic and spintronic applications. \cite{yan2012topological,yan2017topological,armitage2018weyl,he2022topological,vsmejkal2018topological}
In particular, for magnetic materials with broken time-reversal symmetry, anomalous Hall conductivity (AHC) and anomalous Nernst conductivity (ANC) are the most representative linear response transport properties of the topological origin,~\cite{pugh1953hall,nagaosa2010anomalous} making such materials applicable as field sensors, memory devices, thermo-electric power generators and heat-flux sensors.~\cite{gerber2002extraordinary,sakuraba2016potential,zhou2020heat} 
AHC describes the generation of a transverse voltage by an applied longitudinal current, whereas finite ANC emerges under a temperature gradient instead of a current.~\cite{sakuraba2016potential} 
For instance, sufficiently large ANC can be applied to design transverse thermoelectric devices,~\cite{uchida2022thermoelectrics}
thus it is interesting to screen for more candidates exhibiting large AHC and ANC. 

Heusler compounds are a class of materials hosting multifunctional properties driven by their versatile compositions for tunable physical properties. \cite{graf2011simple} They serve as an ideal playground for spintronics because of their high spin polarization and low magnetic damping coefficient.~\cite{hu2020spin} 
In terms of topological properties, there are many ternary half-Heusler compounds which can be classified as topological insulators.~\cite{chadov2010tunable}
Moreover, as demonstrated by Manna {\it et al.}, independent of the magnetization, AHC in the magnetic Heusler compounds can be controlled via the interplay of the crystal symmetry with the topological and geometrical properties of the Berry curvature (BC).~\cite{manna2018colossal} 
Recently, magnetic materials with vanishing band gaps but nontrivial topological properties are becoming interesting, such as spin gapless semiconductors \cite{ouardi2013realization,gao2019high}
and Weyl semimetals.~\cite{manna2018heusler}
It has been reported that the ferromagnetic (FM) Co$_2$MnGa possesses a giant ANC,~\cite{reichlova2018large,sakai2018giant} as well as a strong AHC response.~\cite{belopolski2019discovery} 
Based on the angle-resolved photoemission spectroscopy (ARPES) and the  \textit{ab} \textit{initio} calculations, Belopolski {\it et al.}~\cite{belopolski2019discovery} suggested the potent contribution from topological nodal lines to the large ANC of Co$_2$MnGa. 
However, for Co-based Heulsers, the Seebeck coefficient corresponding to the anomalous Nernst effect (ANE), being 6-8 $\mu$V K$^{-1}$,~\cite{guin2019anomalous,sakuraba2020giant} is still several orders of magnitude smaller than those of nowadays thermoelectric materials. Another route to achieve well-performed transverse thermoelectric properties is via engineering the contributions of the Seebeck effect (SE) and the anomalous Hall effect (AHE). For instance, the Co$_2$MnGa/Si hybrid material hosts a SE-driven transverse thermopower several orders of magnitude higher than the ANE-driven thermopower.~\cite{zhou2021seebeck} In principle, the SE of the thermoelectric material Si generates a longitudinal electric field $ E $ under an applied temperature gradient $\Delta T$, which in turn induces a charge current in the magnetic material Co$_2$MnGa that is then converted into transverse $ E $ by the AHE. 

To search for more Heusler compounds with enhanced topological transport properties, high-throughput (HTP) density functional theory (DFT) calculations are essential, where 
the intrinsic topological properties can be straightforwardly evaluated by constructing accurate tight-binding representation of the electronic structure based on Wannier functions.~\cite{zhang2021high,samathrakis2021enhanced}
Noky {\it et al.}~\cite{noky2020giant} have performed calculations on the intrinsic topological transport properties of the magnetic cubic full Heusler compounds and identified several new Heusler compounds with enhanced AHC and ANC\cite{ji2022spin}. 
In addition, by collecting the experimentally available full and inverse Heusler compounds and evaluating both the spin Hall conductivity (SHC) and AHC, Ji {\it et al.}~\cite{ji2022spin} have suggested such compounds as promising candidates for spintronic applications.
Going beyond the conventional Heusler compounds X$_2$YZ, which are stabilized by the $p$–$d$ covalent bonding between the X and Z atoms, Wei {\it et al.} synthesized the first Hesuler phase Ni$_2$MnTi consisting of only $d$ elements.~\cite{wei2015realization} 
Such all-$d$-metal Heuslers are expected to have very useful mechanical properties due to the $d$–$d$ bonding and have been largely studied in terms of magnetocaloric properties.~\cite{taubel2020tailoring,aznar2019giant,han2019prediction,han2019competition,de2021all} 
Moreover, many compositions have been predicted to show robust magnetic properties, {\it e.g.}, high Curie temperature, tuneable hysteresis properties in useful temperature ranges and capability of maintaining magnetic order upon chemical disorder,~\cite{de2021all,ozdougan2019high} which makes them functional as spintronics.
Note that there has been an HTP screening performed to screen for all-$d$-metal Heuslers, focusing on the thermodynamic stability and the elementary magnetic properties.~\cite{sanvito2017accelerated} However, proper characterization of their topological properties is still missing.

In this work, we conducted HTP calculations to evaluate the AHC and ANC for a total number of 344 ferro/ferri-magnetic all-$d$-metal cubic and tetragonal regular Heusler compounds. 
Compounds with enhanced AHC and ANC are predicted. Moreover, detailed analyses on two typical systems, cubic Re$_2$TaMn and tetragonal Ni$_2$VMn, reveal that the mechanisms for the noticeable AHC and ANC can be attributed to the occurrence of Weyl nodes and the spin-orbit coupling (SOC) induced small gaps close to the Fermi energy, respectively. 

\section{Results and discussion}
\begin{figure}[h!]
	\centering
	\includegraphics[width=1.0\linewidth]{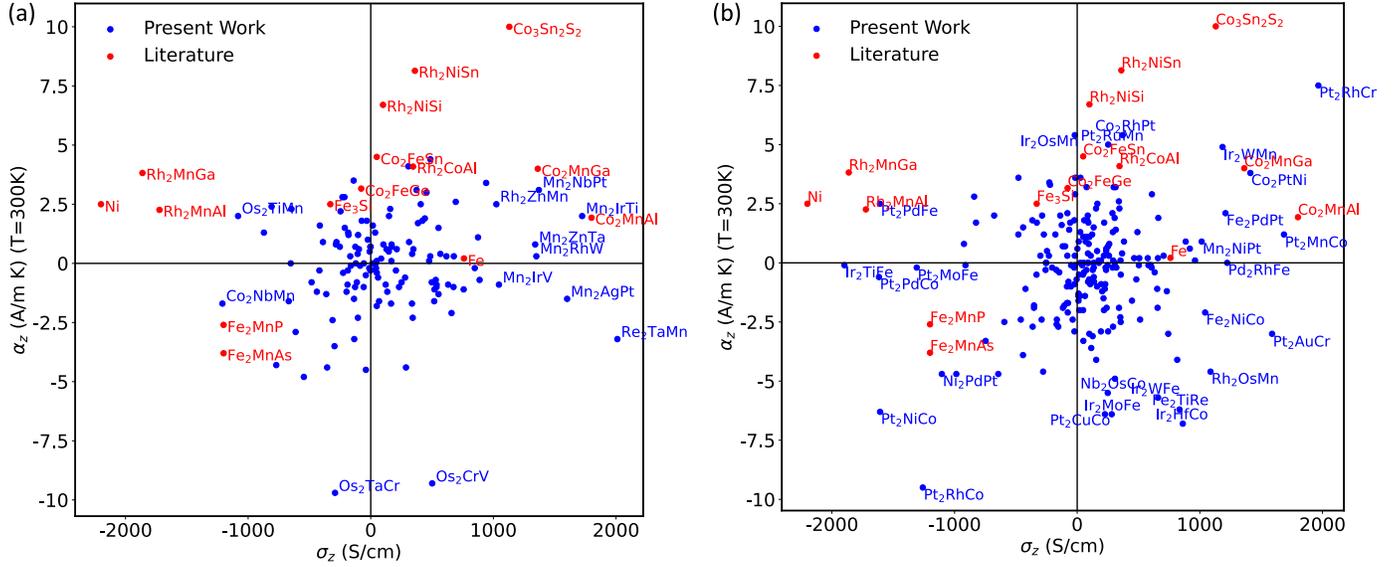}
	\caption{AHC and ANC (at 300 K) of selected materials with the magnetization direction parallel to the [001] direction for (a) cubic and (b) tetragonal all-\textit{d} Heusler compounds. Results of present work are indicated with blue circles and the compositions are displayed for those with AHC larger than 1000 S/cm or ANC larger than 5 A/mK. For comparison, the materials and their AHC/ANC reported in the literature are marked by red circles.}
	\label{figure:ahc_anc}
\end{figure}

The magnetic ground states of the selected 344 all-$d$ Heuslers are confirmed to be FM or ferrimagnetic through our DFT calculations,~\cite{Nuno} thus for the calculations of AHC and ANC, we set the magnetization direction along the [001] axis for both cubic and tetragonal structures.
Among the 344 compounds, 146 of them have cubic structure (Space Group 225) and 198 of them have tetragonal structure (Space Group 139). 
We notice that some of the predicted all-$d$ Heuslers have been successfully synthesized. For instance, Co$_2$MnTi and Mn$_2$PtPd were prepared  by arc melting and their magnetic properties were characterized.~\cite{sanvito2017accelerated} The optimized lattice structure of Co$_2$MnTi in our work is cubic with lattice parameter of 5.84 \AA~(see Table S1), which is in close agreement with the experimental lattice parameter of 5.89 \AA. For Mn$_2$PtPd, experimentally it was found to be antiferromagnetic (AFM) and tetragonally distorted. However, the HTP screening for high AHC and ANC only accounts for FM state in the present work since the collinear AFM structure of Mn$_2$PtPd will not result in finite AHC owing to the presence of inversion symmetry. Recently, Co$_2$MnV was synthesized using the arc-melting method and found to be a cubic Heusler but with multi-phases.~\cite{yu2021tunable} 
Based on the calculated Bain path (see SM Figure S1), a double-well feature at $c$/$a$ = 1 and $c$/$a$ = 1.42 can be observed. Although the Co$_2$MnV is predicted to be more stable under $c$/$a$ = 1.42, the small energy difference between the cubic and tetragonal phases can be readily overcome by, e.g., the temperature effect, which justifies the experimental characterization of being cubic. The DFT-based calculations are performed at zero Kelvin while in reality the synthesis conditions are usually under finite temperatures. Nevertheless, considering the high versatility of available synthesis techniques and processing parameters,~\cite{garmroudi2023unveiling} the DFT-based predictions on the (meta-)stable phases in this work can be regarded as good references to guide future experimental studies.

The calculated $ z $-components of the AHC ($\sigma_z\overset{def}{=}\sigma_{xy}$) and ANC ($\alpha_z\overset{def}{=}\alpha_{xy}$, at 300 K) for the cubic and tetragonal Heuslers are demonstrated in \textbf{Figure~\ref{figure:ahc_anc} } (a) and (b), respectively, 
in comparison with the values of AHC and ANC in the literature.
For the Heuslers with superior transport properties, we list in \textbf{Table~\ref{table:compounds}} the magnitudes of their AHC and ANC.
As listed in Table~\ref{table:compounds}, there are 27 compounds (10 cubic + 17 tetragonal) which show AHC larger than $|1000|$ S/cm and 14 compounds (2 cubic + 12 tetragonal) which exhibit ANC greater than $|5|$ A/mK. 
In particular, three Pt-based tetragonal Heuslers, {\it i.e.}, Pt$_2$RhCr, Pt$_2$NiCo and Pt$_2$RhCo, display simultaneously high AHC and ANC.
Moreover, Re$_2$TaMn is predicted to possess a giant AHC of approximately 2011 S/cm at the Fermi level, which is higher than the already experimentally realized 1600 S/cm for Co$_2$MnGa and 1800-2000 S/cm for Co$_2$MnAl.~\cite{mannatopological,tung2013high}
Additionally, as the maxima of AHC and ANC can be tailored by shifting the Fermi energy level via doping,~\cite{singh2014tuning} Table~\ref{table:compounds} includes also those compounds with significant maximal AHC and ANC (AHC$_{max}$ and ANC$_{max}$) and the corresponding shifts with respect to the Fermi level.
Taking Co$_2$NbMn and Fe$_2$NiCo as examples, their AHC$_{max}$ are -1317 and 1636 S/cm, respectively, which are located at around 0.010 and -0.110 eV with respect to the Fermi level.
From Table~\ref{table:compounds} we can also see that Os$_2$TaCr exhibits the highest magnitude of ANC of -9.7 A/mK, followed by Pt$_2$RhCo (-9.5 A/mK) and Os$_2$CrV (-8.3 A/mK). 
The magnitudes of ANC in these compounds are comparable to the highest ANC reported hitherto of Co$_3$Sn$_2$S (10 A/mK at 70 K).~\cite{yang2020giant} 
Again, despite the relatively low ANC of Co$_2$NbMn and Fe$_2$NiCo at the Fermi level, if doped they could be potential candidates for thermoelectric applications due to the reasonably high ANC$_{max}$ close to the Fermi energy~\cite{zhou2021seebeck,zhou2022seebeck}. 
The full list of the calculated AHC and ANC of the 344 all-$ d $ Heulsers is given in supplementary material (SM) Table S1.

\begin{table}[h!]
	\centering
	\caption{\label{table:compounds}List of Heusler materials (X$_2$YZ ) with promising transport properties, displaying theoretical lattice constants ($a$ and $c/a$), Space Group (SG), AHC and ANC values (at Fermi-level), maximum AHC and ANC values within $\in$ [-0.250,0.250] eV with respect to the Fermi-level ($\Delta E$) and the total magnetic moment ($M_{tot}$) in unit of $\mu_B$ of the primitive cell.}
	\begin{tabular}{l|c|c|c|r|r|r|r|c}
		\hline
		\hline
		Material & $a$ & $c/a$ & SG & AHC & ANC & AHC$_{max}$ ($\Delta E$)& ANC$_{max}$ ($\Delta E$) & $M_{tot}$ \\
		& \AA &  &  & S/cm & A/mK & S/cm (eV) & A/mK (eV) & $\mu_B$ \\
		\hline
		Co$_2$NbMn & 5.96 & 1.00 & 225 & -1210 & -1.7 & -1317 (0.010) & -5.6 (-0.050) & 5.9 \\
		Co$_2$PtNi  & 5.14  & 1.45  & 139  &  1412  & 3.8  & 1646 (0.020)  & -8.9 (0.250)  & 5.0 \\ Co$_2$RhPt & 5.37 & 1.36 & 139 & 371 & 5.4 & -837 (-0.150) & -6.8 (-0.190) & 4.9 \\
		Fe$_2$NiCo  & 5.13  & 1.34  & 139  & 1043  & -2.1  & 1636 (-0.110)  & 6.4 (-0.190)  & 7.6 \\
		Fe$_2$PdPt  & 5.46  & 1.37  & 139  &  1209  & 2.1  & 1423 (0.110)  & -4.8 (0.220) &6.8 \\ 
		Fe$_2$TiRe & 5.95 & 1.00 & 225 & 834 & -6.2 & 1606 (-0.190) & -7.6 (0.030) & 2.8 \\
		Ir$_2$HfCo & 5.60 & 1.38 & 139 & 861 & -6.8 & 1636 (-0.040) & 11.4 (-0.230) & 1.7 \\
		Ir$_2$MoFe & 5.47 & 1.39 & 139 & 282 & -6.4 & 1867 (-0.080) & -7.1 (-0.020) & 2.2 \\
		Ir$_2$OsMn  & 5.38 & 1.42 & 139 & -20 & 5.4 & -1067 (-0.100) & -6.9 (-0.250) & 1.9 \\
		Ir$_2$TiFe & 5.68 & 1.25 & 139 & -1898 & -0.1 & -1931 (-0.010) & 6.6 (0.120) & 3.5 \\
		Ir$_2$WFe &  5.48 & 1.38 & 139 & 660 & -5.7 & 1341 (-0.170) & -5.8 (0.010) & 2.3 \\
		Ir$_2$WMn & 5.48 & 1.38 & 139 & 1185 & 4.9 & 2166 (0.250) & 5.1 (0.020) & 1.8 \\
		Mn$_2$AgPt & 6.33  & 1.00  & 225  &  1601  & -1.5  & 1846 (-0.100)  & 3.6 (-0.240)  & 8.2 \\   
		Mn$_2$IrTi & 6.03  & 1.00  & 225  & 1723  & 2.0  & 1979 (0.060)  & 11.0 (-0.190)  & 4.4 \\
		Mn$_2$IrV  & 5.99  & 1.00  & 225  &  1045  & -0.9  & 1630 (-0.190)  & 5.8 (-0.250)  & 4.1 \\
		Mn$_2$NbPt  & 6.16  & 1.00  & 225  &  1371  & 3.1  & 1482 (0.040)  & -4.7 (0.200)  & 5.1 \\
		Mn$_2$NiPt & 5.77  & 1.16  & 139  &  1015  & 0.9  & 1587 (0.110)  & 2.4 (0.060)  & 8.4 \\
		Mn$_2$RhCo & 6.04 & 1.00 & 225 & 1350 & 0.3 & 1354 (0.004) & -4.2 (0.210) & 5.0 \\
		Mn$_2$ZnTa & 6.02 & 1.00 & 225 & 1341 & 0.8 & 1481 (0.040) & 5.4 (-0.120) & 2.9 \\
		Nb$_2$OsCo & 6.01 & 1.14 & 139 & 249 & -5.5 & 1120 (-0.130) & -5.6 (0.010) &  1.2 \\
		Ni$_2$PdPt & 5.43 & 1.33 & 139 & -1103 & -4.7 & -2675 (0.220) & -9.1 (-0.250) & 2.2 \\
		Os$_2$CrV & 6.03 & 1.00 & 225 & 501 & -9.3 & -1754 (0.150) & -11.4 (0.030) & 2.6 \\
		Os$_2$TaCr  & 6.23  & 1.00  & 225 & -291  & -9.7  & -1401 (0.100)  & -9.7 (0.000)  & 2.8 \\ 
		Os$_2$TiMn  & 6.10  & 1.00  & 225  & -1082  & 2.0  & -1991 (-0.110)  & -7.7 (-0.180)  & 3.0 \\ 
		Pd$_2$RhFe  & 5.42  & 1.45  & 139  & 1223  & 0.0  & 1354 (-0.010)  & -6.7 (0.240)& 5.1 \\
		Pt$_2$AuCr & 5.64 & 1.40 & 139 & 1589 & -3.1 & 2040 (-0.030) & 5.4 (-0.190) &2.7 \\		
		Pt$_2$CuCo & 5.47 & 1.34 & 139 & 226 & -6.4 & 1342 (-0.090) & -6.5 (0.010) & 2.5 \\
		Pt$_2$MnCo & 5.47 & 1.38 & 139 & 1686 & 1.2 & 1839 (0.030) & 7.2 (-0.250) & 5.9 \\
		Pt$_2$MoFe & 5.42 & 1.47 & 139 & -1308 & -0.2 & -1345 (0.010) & -3.4 (-0.060) & 2.2 \\
		Pt$_2$NiCo & 5.41 & 1.36 & 139 & -1607 & -6.3 & -2158 (0.030) & -7.6 (0.030) & 3.5 \\		
		Pt$_2$PdCo & 5.47 & 1.42 & 139 & -1616 & -0.6 & -1715 (0.030) & -8.8 (-0.240) & 3.0 \\
		Pt$_2$PdFe & 5.53 & 1.41 & 139 & -1608 & 2.5 & -1694 (-0.180) & 9.2 (0.080) & 4.0 \\
		Pt$_2$RhCo & 5.43  & 1.42  & 139  & -1259  & -9.5 & -2154 (0.040)  & -10.2 (-0.020)& 3.1\\ 
		Pt$_2$RhCr & 5.46 & 1.43 & 139 & 1966  & 7.5 & 2432 (0.050) & 10.1 (-0.050) & 1.4 \\
		Pt$_2$RuMn & 5.43 & 1.44 & 139 & 253 & 5.0 & 1154 (0.120) & -5.4 (0.150) & 2.8 \\
		Re$_2$TaMn & 6.23 & 1.00 & 225 & 2011 & -3.2 & 2031 (0.003) & -8.2 (0.060) & 2.3 \\
		Rh$_2$OsMn  & 5.35  & 1.43 & 139  & 1087  & -4.6  & 2113 (-0.080)  & -5.8 (0.160) &2.3 \\ 		
		Rh$_2$ZnMn  & 6.03  & 1.00  & 225  &  1024  & 2.5  & 1959 (0.230)  & -5.1 (0.250)  & 3.2 \\   
		\hline
	\end{tabular}
\end{table}

\begin{figure}[h!]
	\centering
	\includegraphics[width=\linewidth]{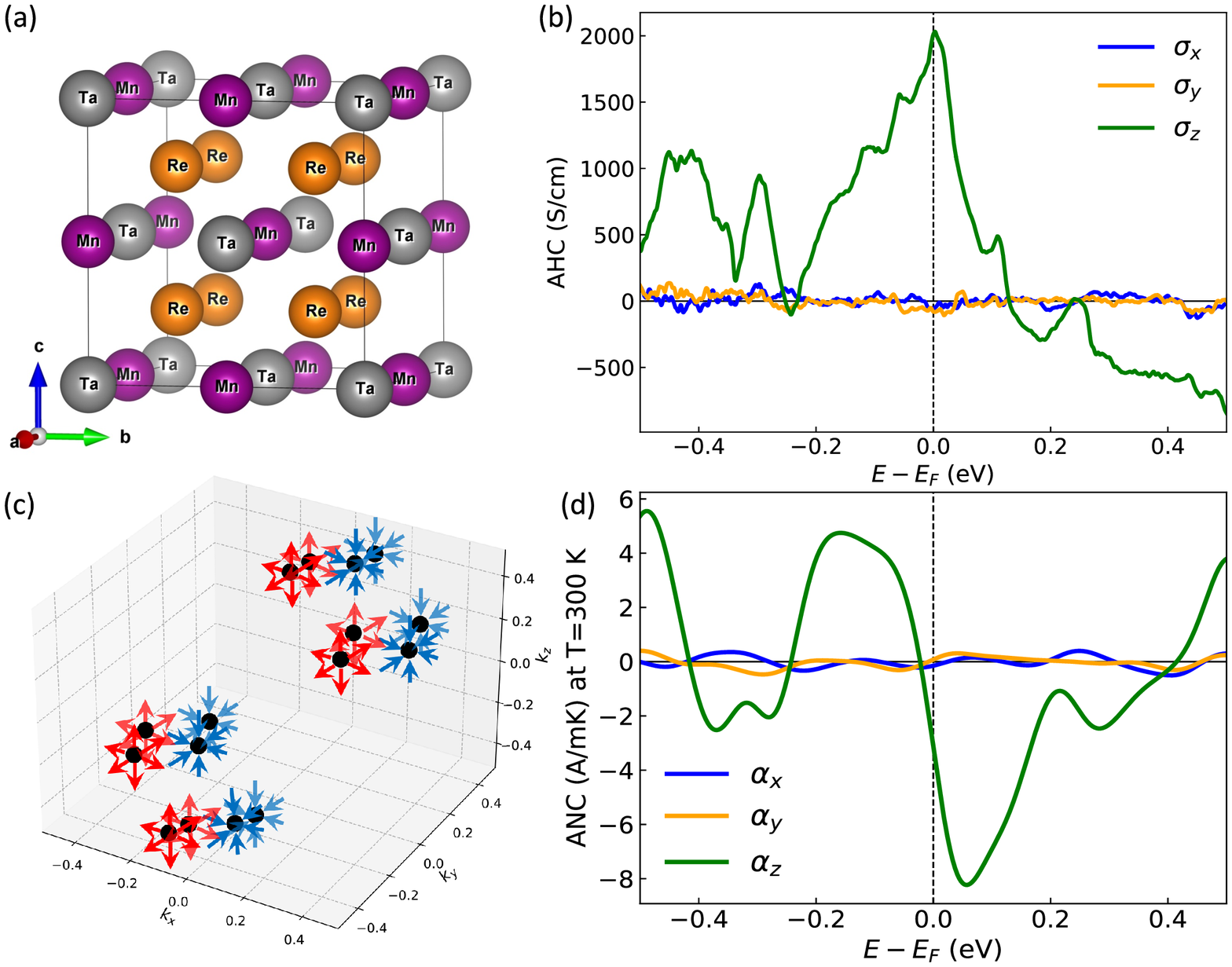}
	\caption{(a) Crystal structure of Re$_2$TaMn (conventional cell). (b) The components of the calculated AHC, $\sigma_x$, $\sigma_y$ and $\sigma_z$ of Re$_2$TaMn, for the magnetization direction along the [001] direction. (c) Illustration of the 16 identified Weyl points (in the k space with the unit of $2\pi/a$) in the Brillouin zone with Berry flux direction (The calculation is adopted in the primitive cell). Red and blue arrows around the Weyl point denote positive (source) and negative (sink) topological charges, respectively. (d) The components of the calculated ANC at 300 K of Re$_2$TaMn, $\alpha_x$, $\alpha_y$ and $\alpha_z$, for the magnetization direction along the [001] direction.
	}
	\label{figure:Re2}
\end{figure}

To gain insights on the origin of the giant AHC and ANC, detailed analysis has been conducted on the electronic structure for two representative cases, {\it i.e.}, cubic Re$_2$TaMn and tetragonal Ni$_2$VMn. 
The crystal structure of Re$_2$TaMn is displayed in \textbf{Figure~\ref{figure:Re2}} (a) and the magnetization direction is along the [001] direction. The calculated AHC and ANC curves with respect to the Fermi energy are displayed in Figure~\ref{figure:Re2} (b) and (d), respectively. An anomalous peak can be observed at the Fermi level for the AHC with the peak value being approximately 2011 S/cm. Regarding the ANC at 300 K, the maximum is located at 0.060 eV above the Fermi level and shows an absolute value as high as 8.2 A/mK. Additionally, it can be noticed that the $ x $- and $ y $- components of AHC and ANC have vanishing magnitudes, suggesting that the AHC and ANC vector is parallel to the magnetization direction, as it can be interpreted based on symmetry. \cite{zhang2011anisotropic}

Specifically, the shape of the AHC tensor depends on the direction of the magnetization. Since BC transforms as a pseudovector under the symmetry operations,~\cite{suzuki2017cluster} the AHC and ANC tensors can be determined by finding the transformation of BC for all symmetries of the magnetic Laue group according to:
\begin{equation}\label{BCVECTOR}
	s\boldsymbol{\Omega}(\boldsymbol{r})=\pm \det (\boldsymbol{D}(R))D(R)\boldsymbol{\Omega}(s^{-1}\boldsymbol{r}),
\end{equation}
where the BC pseudovector is represented by $\boldsymbol{\Omega}(r)$, the three-dimensional representation of a symmetry operation (excluding the translation symmetry) is given by $\boldsymbol{D}(R)$, and an arbitrary symmetry operation is denoted by $s$. 
As a matter of fact, all the investigated Heusler compounds in the current work (both cubic and tetragonal) fall into the magnetic Laue group 4/mm$^\prime$m$^\prime$ when the magnetization is applied along the [001] direction. 
In this case, the m$_z$ mirror symmetry transforms the three components of the BC pseudovector in the following manner:
\begin{align}
	& \Omega_x(k_x, k_y, -k_z) = -\Omega_x(k_x, k_y, k_z), \notag\\\
	& \Omega_y(k_x, k_y, -k_z) = -\Omega_y(k_x, k_y, k_z), \notag\\\
	& \Omega_z(k_x, k_y, -k_z) = \Omega_z(k_x, k_y, k_z). 
\end{align}
Therefore, the summation of the BC over the entire Brillouin zone forces the AHC along the $ x $- and $ y $-axis to be zero, while leaving the component along the $ z $-axis unrestricted, as the AHC itself depends on the BC according to Eq.~\eqref{AHC}. 

\begin{figure}[h!]
	\centering
	\includegraphics[width=0.9\linewidth]{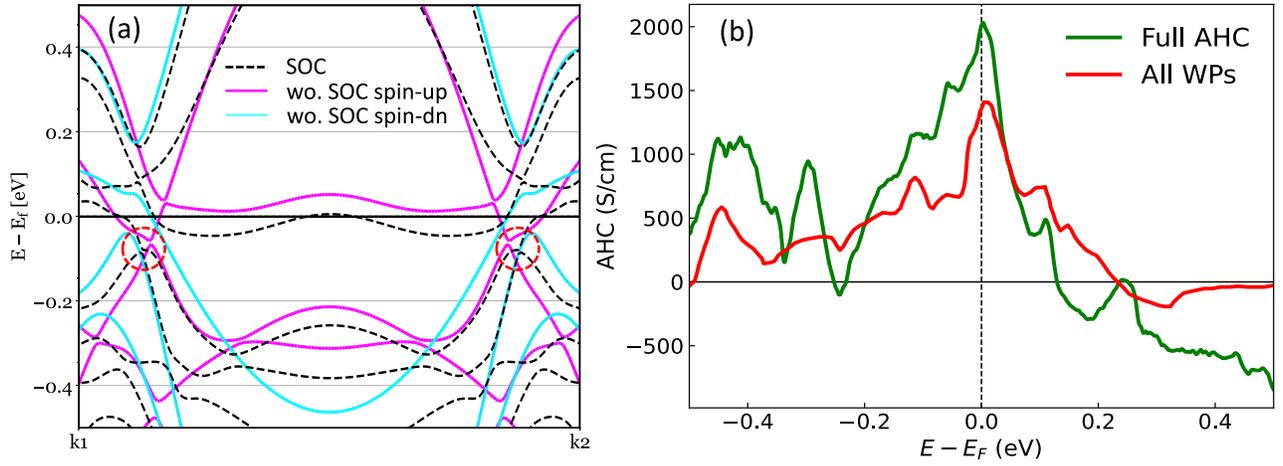}
	\caption{(a) The band structure of primitive Re$_2$TaMn along the $ k $ path from k1 (-0.10, 0.35, 0.41) to k2 (0.35, -0.10, 0.41) which crosses a pair of Weyl nodes (marked out by red dashed circles) with spin-orbit coupling (SOC) effect included. The band structure in the spin-polarized case without SOC is shown for comparison (b) The z-component of the AHC ($\sigma_z$) (green) and the contribution from all Weyl points to AHC (red) as a function of energy (with respect to the Fermi level) for Re$_2$TaMn with magnetization along the [001] direction.}
	\label{figure:contribution}
\end{figure}

Furthermore, as Weyl points act as sources or sinks of BC~\cite{berry1984quantal,vanderbilt2018berry,xiao2010berry} which can lead to significant AHC and ANC when they are close to the Fermi energy,~\cite{samathrakis2021enhanced,singh2021multifunctional}
we searched for Weyl points within $\pm$50 meV around the Fermi energy. 
A total of 16 symmetry-related Weyl points were identified in Re$_2$TaMn, as illustrated in Figure~\ref{figure:Re2}~(c) with the positive (negative) chiral charges marked by red (blue) arrows indicating the spatial direction of the Berry flux.
The chirality of each Weyl point was calculated in terms of the shift of the Wannier charge centres~\cite{vanderbilt2018berry} and we list in \textbf{Table~\ref{table:chira}} the symmetry operations in the magnetic Laue group 4/mm$^\prime$m$^\prime$, as well as their relations with the chirality. It can be first noticed that the sum of the chirality of the Weyl points within the Brillouin Zone (BZ) vanishes, as demonstrated in Nielsen-Ninomiya's theorem of Fermion doubling.~\cite{nielsen1981absence}
Additionally, it can be explicitly seen from Table~\ref{table:chira} that the mirror planes, inversion symmetry and roto-inversions are responsible for the sign change of chirality.
In \textbf{Figure~\ref{figure:contribution}} (a), we plot out the band structure with SOC included along a specific $ k $ path which crosses a pair of Weyl nodes (-0.04, 0.30, 0.41) and (0.30, -0.04, 0.41). The presence of the Weyl points can be clearly observed, which are located at approximately 35 meV below the Fermi level. The calculated band structure without SOC effect is shown in Figure~\ref{figure:contribution} (a) for comparison. The Weyl point feature exists in the spin-up channel. 
The SOC effect, which mixes the spin-up and spin-down channels, slightly alters the $ k $ coordinates of the Weyl pair with respect to the spin-polarized case without SOC, and the energy position corresponding to the Weyl pair is closer to the Fermi level. 
The contribution of each Weyl point to the total AHC was subsequently evaluated by calculating the AHC within cubes of 0.17 unit length (in the reciprocal space) while keeping the Weyl points at the cubic centre. 
The total contributions of all the 16 Weyl points, whose total volume amounts to about 7.8\% of the whole BZ, are then obtained by summing up the AHC resulting from the 16 cubes, which constitutes approximately 74\% of the total AHC at the Fermi level (see Figure~\ref{figure:contribution}~(b)).

\begin{table}[h!]
	\centering
	\caption{\label{table:chira} Symmetry operations of the magnetic Laue group 4/mm$^\prime$m$^\prime$ and the relation with chirality. Here (x,y,z) form includes the time reversal symmetry.}
	\begin{tabular}{l|c|c|c|c}
		\hline
		\hline
		N & (x,y,z) form & TR & Seitz Symbol & Chirality \\
		\hline
		1 & $x$, $y$, $z$ & 1 & 1 & 1 \\
		2 & $-x$, $-y$, $z$  & 1 & $2_z$ & 1 \\
		3 & $-y$, $x$, $z$  & 1 & $4_z$ & 1 \\
		4 & $y$, $-x$, $z$ & 1 & $4_z^{-1}$ & 1 \\
		5 & $-x$, $-y$, $-z$  & 1 & $-1$ & -1 \\
		6 & $x$, $y$, $-z$  & 1 & $m_z$ & -1 \\
		7 & $y$, $-x$, $-z$  & 1 & $-4_z$ & -1 \\
		8 & $-y$, $x$, $-z$ & 1 & $-4_z^{-1}$ & -1 \\
		9 & $-x$, $y$, $z$  & -1 & $2_x^{\prime}$ & 1 \\
		10 & $x$, $-y$, $z$  & -1 & $2_y^{\prime}$ & 1 \\
		11 & $y$, $x$, $z$  & -1 & $2_{-xy}^{\prime}$ & 1 \\
		12 & $-y$, $-x$, $z$  & -1 & $2_{xy}^{\prime}$ & 1 \\
		13 & $x$, $-y$, $-z$  & -1 & $m_{x}^{\prime}$ & -1 \\
		14 & $-x$, $y$, $-z$  & -1 & $m_{y}^{\prime}$ & -1 \\
		15 & $-y$, $-x$, $-z$  & -1 & $m_{-xy}^{\prime}$ & -1 \\
		16 & $y$, $x$, $-z$ & -1 & $m_{xy}^{\prime}$ & -1 \\
		\hline
	\end{tabular}
\end{table}

\begin{figure}[h!]
	\centering
	\includegraphics[width=\linewidth]{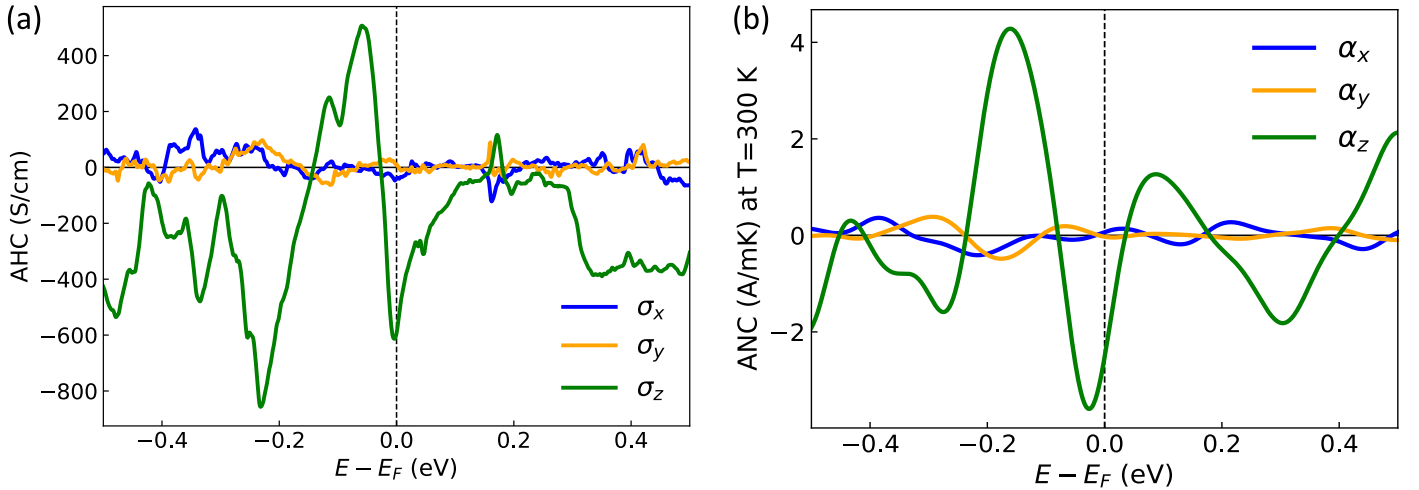}
	\caption{(a) The components of the calculated AHC, $\sigma_x$, $\sigma_y$ and $\sigma_z$, and (b) the calculated ANC at 300 K, $\alpha_x$, $\alpha_y$ and $\alpha_z$, of Ni$_2$VMn for the magnetization direction along the [001] direction with respect to the Fermi energy. }
	\label{figure:Ni2MnV}
\end{figure}

\begin{figure}[h!]
	\centering
	\includegraphics[width=\linewidth]{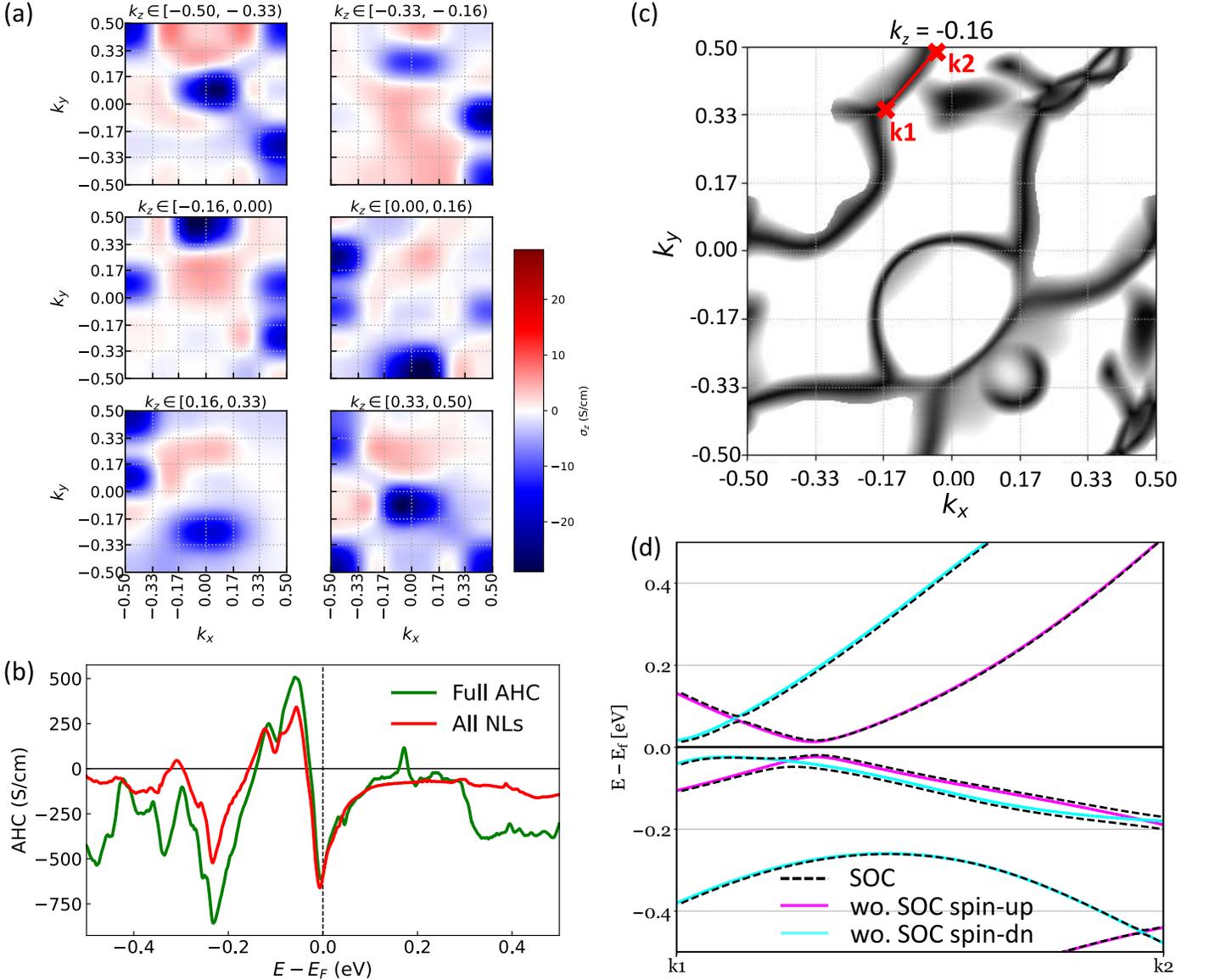}
	\caption{(a) The z-component of the AHC evaluated in 216 Brillouin Zone cubes for Ni$_2$VMn with magnetization direction parallel to the [001] axis.
		(b) The z-component of the AHC ($\sigma_z$) (green) and the contribution from the nodal lines (NLs) to AHC (red) as a function of energy for Ni$_2$VMn with magnetization along the [001] direction. (c) Small band gap area between bands in the vicinity of Fermi level at a slice of Brillouin zone of $k_z=-0.16$. (d) The calculated band structure with spin-orbit coupling (SOC) effect along the path from k1 to k2 is compared with that of spin-polarized calculation without SOC effect.}
	\label{figure:nodal_contribution}
\end{figure}

In contrast to Re$_2$TaMn, we find that the singular behaviour of the AHC and ANC in Ni$_2$VMn can be attributed to a different origin. 
Ni$_2$VMn is a regular tetragonal Heusler compound with $c/a=1.48$ and total spin moment $\mu_{S}=4.16~\mu_B$, in which Ni, Mn and V have magnetic moments of 0.31, 2.40 and -0.92 $\mu_B$ respectively. The AHC and ANC of Ni$_2$VMn are shown in \textbf{Figure~\ref{figure:Ni2MnV}} (a) and (b), respectively. The absolute values of AHC and ANC at the Fermi level are 594.0 S/cm and 2.5 A/mK, respectively. 
In order to identify the main contributor to the AHC, we divided the $ k $ space corresponding to the primitive cell of Ni$_2$VMn within $k_i \in [-0.5,0.5)$ ($i=x,y,z$) into $6\times6\times6$ cubes and calculated the AHC within each cube (see the colour map displayed in \textbf{Figure~\ref{figure:nodal_contribution}} (a)). 
The sum of the AHC values obtained in the 216 cubes is around -552.5 S/cm, which is close but slightly different from the calculated AHC by integrating the whole BZ owing to the different k-mesh density.
It can be seen from Figure~\ref{figure:nodal_contribution} (a) that, the dominant contributions to the AHC come from the blue squared regions. By summing up the AHC values obtained from only the blue areas, we obtain a total AHC of around -592.2 S/cm. Therefore, the peak value of AHC around the Fermi level can be well reproduced by considering only the blue areas in the BZ, whereas the other areas have limited contribution to the total AHC (see Figure~\ref{figure:nodal_contribution} (b)).
 
Additionally, we notice that the main contributing areas are correlated with small band gap areas by highlighting the $ k $ points which possess small band gap. Taking the plane with $k_z=-0.16$ as an example (with reference to the middle left panel of Figure~\ref{figure:nodal_contribution} (a)), the shaded areas in Figure~\ref{figure:nodal_contribution} (c) are with band gaps smaller than 27 meV between two neighbouring bands in the vicinity of the Fermi level. It can be found that for the region with $k_x \in (-0.17,0,17)$ and $k_y \in (0.33,0,50)$, the relatively large component of $\sigma_z$ is manifested in small band gap areas. We then selected a specific $ k $ path on this $ k $ plane, \textit{i.e.}, from k1 to k2 which are marked out by the cross symbols, to investigate how the band structure changes by including the SOC effect. It can be clearly seen from Figure~\ref{figure:nodal_contribution} (d) that the band crossings between the spin-up and spin-down channels in the spin-polarized calculation without SOC effect are avoided when SOC effect is considered, resulting in the continuous small band gap opening slightly below the Fermi level along the selected $ k $ path. Therefore, the singular AHC and ANC in Ni$_2$VMn are mainly driven by a gap opening of nodal lines due to SOC. 

\begin{figure}[h!]
	\centering
	\includegraphics[width=0.8\linewidth]{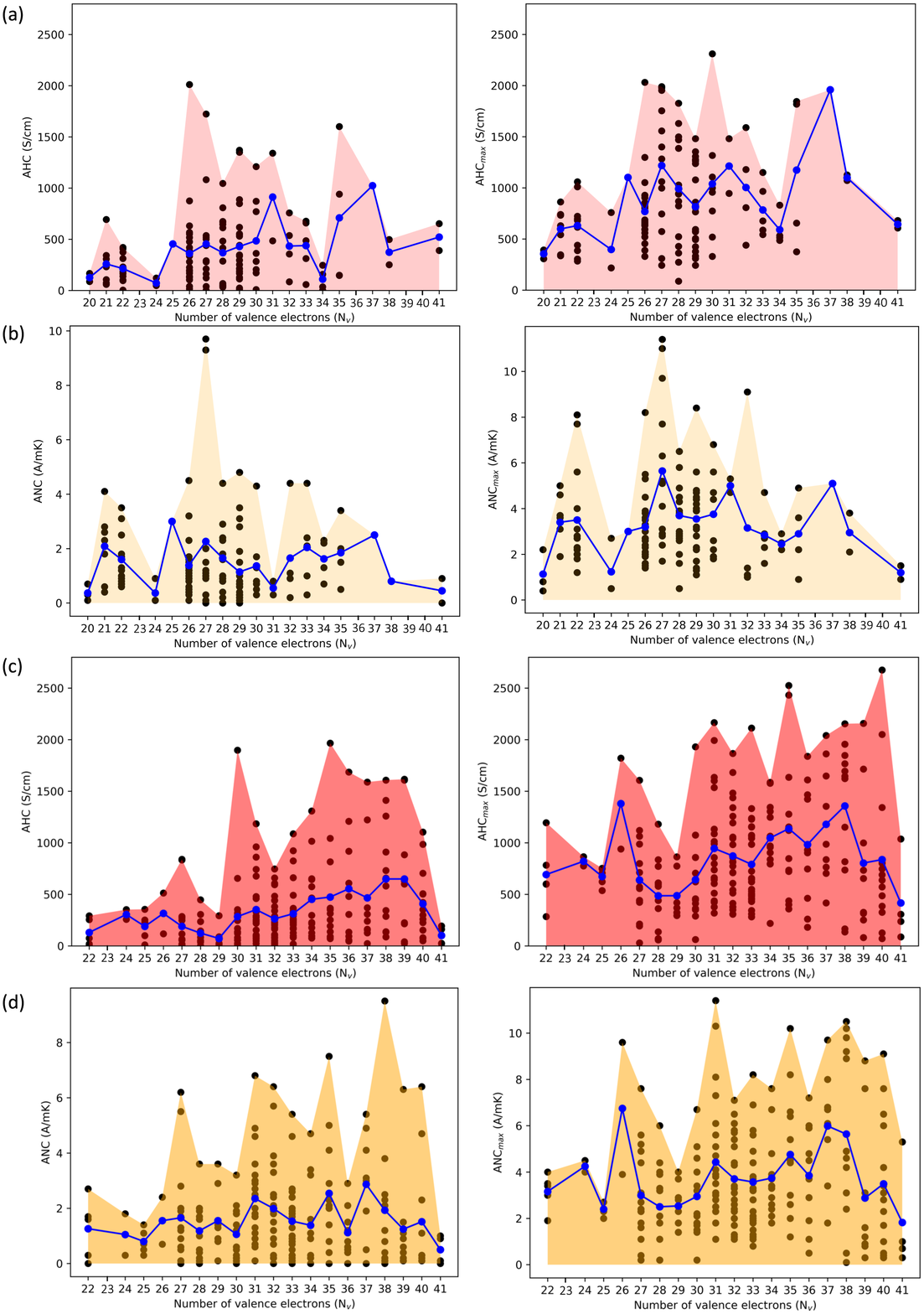}
	\caption{(a) The AHC at Fermi level, AHC$_{max}$, (b) ANC at Fermi level, ANC$_{max}$ at 300 K with respect to the number of valence electrons ($N_v$) for regular cubic all-\textit{d} Heuslers; (c) the AHC at Fermi level, AHC$_{max}$, (d) ANC at Fermi level, ANC$_{max}$ at 300 K with respect to the number of valence electrons ($N_v$) for regular tetragonal all-\textit{d} Heuslers.}
	\label{figure:ahc_c_nv}
\end{figure}

Based on the current HTP work, it is also interesting to inspect the correlations between the transport properties, \textit{i.e.}, AHC, AHC$_{max}$, ANC, ANC$_{max}$, and the number of valence electrons ($N_v$). We inspect such correlations separately in cubic and tetragonal Heuslers, as displayed in \textbf{Figure~\ref{figure:ahc_c_nv}} (a) to (d), respectively. We find that for regular cubic Heuslers, the peak for both AHC and ANC usually occurs for N$_v \in [26,28]$, whereas the summits are usually seen for $N_v > 30$, particularly for $N_v \in [36,40]$, in tetragonal Heuslers. Such correlations can be understood from the point that the number of valence electrons is related to the band filling of the band structure. Namely, certain valence electron counting can move the Fermi level closer to the topological features, thus leading to higher AHC and ANC. Clearly the band filling behaves differently in cubic and tetragonal structures. However, it should be reminded that large AHC does not necessarily means large ANC, as pointed out in Ref.~\cite{noky2020giant}. We also notice that, despite the different space groups the cubic and tetragonal Heuslers belong to, they actually have the same magnetic Laue group 4/mm$^\prime$m$^\prime$ under the studied ferro/ferri-magnetic state. 
From a symmetry point view, their belonging to the same magnetic Laue group renders the similarity in their performances on the transport properties. By classifying the absolute values of AHC/ANC by the space group, we indeed find that the maxima, as well as the averages of the AHC/ANC for space group No.139 and No.225 are almost equivalent. The maximum and average of AHC in both cubic and tetragonal Heuslers are around 2000 and 370 S/cm, respectively. Similarly, the maximum and average of ANC at 300 K in both cubic and tetragonal Heuslers are about 10.0 and 1.8 A/mK, respectively.

\section{Conclusion}
In summary, we have carried out a HTP screening of all-$d$ regular magnetic Hesulers, including both cubic and tetragonal compounds, to search for Heuslers with large AHC/ANC. A giant AHC of 2011 S/cm at the Fermi level has been found in cubic Re$_2$TaMn, following the tetragonal Pt$_2$RhCr with an AHC of 1966 S/cm. Interestingly, Co$_2$NbMn and Fe$_2$NiCo, which consist of more abundant elements, are predicted to have rather high |AHC| at the Fermi level, being approximately 1210 and 1043 S/cm, respectively. Additionally, Os$_2$TaCr and Pt$_2$RhCo exhibit high magnitudes of |ANC| being 9.7 and 9.5 A/mK, respectively, which are comparable to the highest ANC reported hitherto of Co$_3$Sn$_2$S. The current work also reveals the determining role of symmetry on the topological properties. Belonging to the same magnetic Laue group 4/mm$^\prime$m$^\prime$, the ferro/ferri-magnetic cubic and tetragonal Heuslers display similar performance regarding the maxima and averages of |AHC|/|ANC|. Moreover, we find that the AHC/ANC values are mainly contributed by the Weyl points, avoided band crossing, as well as the nodal lines with small band gaps. 

Furthermore, the exact position of the BC with respect to the Fermi level significantly influences the transport properties. Therefore, we further illustrate separately for cubic and tetragonal Heuslers the correlations between the AHC, AHC, AHC$_{max}$, ANC, ANC$_{max}$ and the number of valence electrons $N_v$. For cubic Heuslers, we have identified $N_v \in [26,28]$ as the sweet points, whereas for tetragonal Heuslers, the compounds with $N_v \in [36,40]$ stand out. Such statistical analysis provides a valuable guidance for tailoring the transport properties of Heuslers via altering the band filling by, \textit{e.g.}, chemical doping, as recently reported for Fe$_3$Sn~\cite{samathrakis2022thermodynamical} and (Ti,Zr,Hf)MnP.~\cite{samathrakis2022tunable}


\section{Experimental Section}
A total of 344 Heusler compounds X$_2$YZ, which have negative formation energies and convex hull distance smaller than 175 meV/atom, were selected for the evaluation of the topological transport properties (AHC and ANC) based on our HTP calculations.~\cite{Nuno} 
For Heuslers X$_2$YZ, X (0.25, 0.25, 0.25), Y (0.00, 0.00, 0.00) and Z (0.50, 0.50, 0.50) are one of the $3d/4d/5d$ elements except Tc, and at least one magnetic $3d$ element (V, Cr, Mn, Fe, Co, Ni) is on either the X or the Y/Z sites.
The selected materials were thermodynamically stable/metastable, FM or ferrimagnetic in nature, and had regular cubic or regular tetragonal ground state structure. 
The calculations were automated by using an in-house developed workflow~\cite{zhang2018high,autowanflow} that integrates VASP, ~\cite{kresse1993ab,kresse1994ab,kresse1996efficiency}
Wannier90~\cite{mostofi2014updated} and WannierTools~\cite{WU2018} software packages. 
Firstly, the DFT calculations were conducted using VASP, in which the projected augmented wave (PAW)~\cite{blochl1994projector,kresse1999ultrasoft} method and the exchange-correlation functional of GGA-PBE~\cite{perdew1996generalized} were implemented. 
All the calculations included the SOC effect.  
The $ k $ mesh used for the calculations, being $\Gamma$-centred, had a density of 50 with respect to the lattice parameter. The cut-off energy for the plane waves was set to 500 eV. 
Subsequently, the maximally localized Wannier functions (MLWFs)~\cite{marzari2012maximally} were constructed using the Wannier90 code. 
The required parameters, such as the number projection orbitals and the disentanglement and frozen windows, were chosen automatically according to the algorithm proposed by Zhang {\it et al.}~\cite{zhang2018high}
Lastly, the intrinsic AHC was evaluated using WannierTools based on the tight binding model constructed from the MLWFs, which is defined as the integration of BC over the entire BZ. Summing over all of the occupied bands, the AHC can be obtained according to the following formula~\cite{wang2006ab}

\begin{equation}\label{AHC}
	\sigma_{\alpha\beta} = -\frac{e^2}{\hslash}\int\frac{d\boldsymbol{k}}{(2\pi)^3}\sum_n f[\epsilon(\boldsymbol{k})-\mu]\Omega_{n,\alpha\beta}(\boldsymbol{k}),
\end{equation}
with BC itself calculated according to 
\begin{equation}
	\Omega_{n,\alpha\beta} (\boldsymbol{k}) =  -2\mathrm{Im}\sum_{m\neq n} \frac{ \langle {\boldsymbol{k}n} \mid {\nu_\alpha (\boldsymbol{k})} \mid {\boldsymbol{k}m} \rangle \langle {\boldsymbol{k}m} \mid {\nu_\beta(\boldsymbol{k})} \mid {\boldsymbol{k}n} \rangle}{\left[\epsilon_m(\boldsymbol{k})-\epsilon_n(\boldsymbol{k})\right]^2},
\end{equation}
in which $\mu$ stands for the Fermi energy, $ n $ and $ m $ for the occupied and unoccupied bands, respectively, and $\epsilon_n(\boldsymbol{k})$ and $\epsilon_m(\boldsymbol{k})$ for the corresponding eigenvalues. $f[\epsilon(\boldsymbol{k})-\mu]$ represents the Fermi distribution function. The $\nu_{\alpha}$ and $\nu_{\beta}$ are velocity operators. 
In order to evaluate the AHC, a $ k $ mesh having the density of 400 with respect to the lattice parameter was used to guarantee good convergence.
The ANC was calculated by evaluating the following integral,~\cite{xiao2006berry}
\begin{equation}\label{ANC}
	a_{\alpha\beta} = -\frac{1}{e} \int d\epsilon \frac{\partial f}{\partial \mu} \sigma_{\alpha\beta}(\epsilon)\frac{\epsilon-\mu}{T},
\end{equation}
where $\epsilon$ represents the energy of a point in the energy grid, $e$ and $T$ are respectively the electron charge and the temperature. The above integral was evaluated using an energy grid containing 1001 points within the range [-0.5,0.5] eV with respect to the Fermi level at 300 K. The detailed analysis on the origin of AHC utilized the available features of WannierTools,~\cite{yu2011equivalent,soluyanov2011wannier} including the search for Weyl or nodal points, and the calculations of chirality, BC and \textit{etc}.

\medskip
\textbf{Supporting Information} \par 
Supporting Information is available from the Wiley Online Library or from the author.

\medskip
\textbf{Acknowledgements} \par 
The Lichtenberg high performance computer of the TU Darmstadt is gratefully acknowledged for the computational resources where the calculations were conducted for this project. This project was supported by the Deutsche Forschungsgemeinschaft (DFG, German Research Foundation)-Project-ID 405553726-TRR 270, and the European Research Council (ERC) under the European Union’s Horizon 2020 research and innovation programme (Grant No. 743116-project Cool Innov)

\medskip

\bibliographystyle{MSP}

\end{document}